# Ground sculpting to enhance vertical bifacial solar farm output

M. Ryyan Khan[1], Enas Sakr[2], Xingshu Sun[2], Peter Bermel[2], and Muhammad A. Alam[2, a)]

[1]Department of Electrical and Electronic Engineering, East West University, Dhaka, Bangladesh
[2]Electrical and Computer Engineering Department, Purdue University, West Lafayette, IN 47907, USA

The prospect of additional energy yield and improved reliability have increased commercial interest in bifacial solar modules. A number of recent publications have quantified the bifacial gain for several configurations. For example, a *standalone*, optimally-tilted bifacial panel placed over a *flat ground* (with 50% albedo) is expected to produce a bifacial energy gain of 30% (per module area). In contrast, for a panel array in a solar farm, self and mutual shading reduce the bifacial gain at the same tilt to 10-15% (per farm area). Bifacial gain is negligible for vertical arrays — although the configuration is of significant interest, since it can prevent soiling. Here, we calculate the bifacial gain of a solar farm where vertical arrays have been placed over sculpted/patterned ground. We conclude that vertical panels straddling (upward) triangle-shaped ground maximizes the energy output. For this optimum configuration, the bifacial gain can approach 50%, especially for regions with moderate to high cloudy conditions. The enhanced output, along with reduced soiling loss and lower cleaning cost of the ground sculpted vertical bifacial (GvBF) solar farm could be of significant technological interest, especially in regions such as the Middle East and North Africa (MENA), particularly susceptible to significant soiling losses.

## 1. Introduction and background

Monofacial panels are the most commonly used panel configuration in today's photovoltaic (PV) industry. Recent trends, however, show a steady increase in the share of bifacial panels in the PV market, and ITRPV also predicts further increase in market share of bifacial PV over the next decade [1]. This prediction is not universally accepted, since bifacial PV is typically more expensive. Unless bifacial power gains can offset additional module costs, the technology would have higher levelized cost of energy (LCOE).

Allowing light to be captured from both faces thereby creating a 'bifacial' panel dates back to the early 1980s [2]–[4]. Cuevas *et al*. [3] showed ~50% gain in bifacial output is possible, compared to a monofacial panel. However, this analysis requires a white painted ground and strategically placed white wall to obtain such high gains. A more recent detailed numerical analysis by Sun *et al*. [5] predicts that the optimally placed bifacial panel over a flat ground will have energy gain of only 10% for albedo of 25% (e.g., concrete). For artificially-treated ground with albedo 50% (e.g., white concrete), the bifacial gains can be 30%. These results agree with various independent calculations published in recent years [6]–[10]. The bifacial gain, therefore, may not be as high as anticipated originally.

The analysis of large solar farms [11]–[14] based on bifacial panel arrays is even more complex due to mutual shading among the panels, as well as periodic shading on the ground [15], [16]. Our recent work [16] showed that, for optimally designed farms, vertical bifacial solar farm produces 10-30% *less* energy compared to an optimum monofacial farm. Although tilting the bifacial module does improve the yield (~10%) [15], the attractive soiling-resistant property of vertical farm is lost. Lower soiling translates into longer cleaning cycles, better integrated energy yield, and lower cleaning cost.

Given the relatively small bifacial gain for vertical farms, one might consider resurrecting the ideas of Cuevas *et al*. (in the context of the solar farms) to improve the energy output of the bifacial solar farms. Recall that Cuevas *et al*. [3] obtained high bifacial gains by increasing the backside reflection from a white vertical wall. This configuration may be viewed as a cleverly-designed low-concentration bifacial PV. Indeed, there have been several designs and studies focusing on low-concentration conventional PV [17]–[19] and bifacial PV systems [20]–[22]. A recent study [22] experimentally demonstrated a comparison between flat and parabolic reflectors for tilted bifacial panels. We should also be able to adapt these ideas of ground-sculpting to vertical bifacial farms.

In this paper, we infuse the idea of low concentration PV with vertical bifacial PV by artificially sculpting the ground to enhance albedo collection on the two faces of the panel. We have developed a model to analyze arbitrarily shaped ground between periodic arrays of vertical panels. The diurnal and seasonal variations of sun path and solar illumination are taken into account using our previously developed model [16]. Our analysis predicts that, among various configurations, ground mounted vertical bifacial PV with upward triangular ground shapes is optimal for maximizing annual energy yield. For an effective albedo reflectance of $R_A \sim 0.5$, we observe that the designed bifacial PV farm always yields higher than a monofacial farm. In fact, at locations with low clearness index (higher diffused light), the bifacial gain is ~50%. While $R_A \sim 50\%$ can be seen for white concrete, artificially designed white roofing foils can have $R_A > 80\%$ [23]. One must balance the additional gain with the additional cost of ground-sculpting to assess the viability of the approach.

---

[a)] Electronic mail: alam@purdue.edu.

## 2. Numerical model.

In our numerical model (see Fig. 2 for a summary), we obtain the irradiance information based on a NASA meteorological database [24]. The irradiance data and the panel array configuration is then used to find the light incident on the panel faces and the ground. Some of the light scattered from the ground is collected by the bifacial panel faces. An electrical model for the panel then calculates the energy output corresponding to the collected sunlight. The hourly output is integrated to predict the annual yield. The details of each of these calculation steps are discussed below.

### 2.1. Solar Data:

The daily average meteorological NASA data [24] is combined with the clear-sky model from Sandia [25] to calculate a time series of insolation information. For any given location on the world, we find minute-by-minute variation of Global Horizontal Irradiance (GHI or $I_{GHI}$), Direct Normal Irradiance (DNI or $I_b$), Diffused Horizontal Irradiance (DHI or $I_{\text{diff}}$), solar azimuth angle $\gamma_S$, and solar zenith angle $\theta_Z$. More detailed description is provided in our prior publications [5], [16].

### 2.2. Panels and array configuration:

In this paper, we consider an array of vertically placed bifacial panels facing East-West (E-W). Conventionally, the ground is kept horizontal and flat. However, intuitively, we expect additional albedo reflection if the ground is sculpted. In general, the panels may be elevated over arbitrarily patterned ground (as shown in Fig. 1 (a)). Since the vertical panel array is periodic, the ground pattern should be periodic (and symmetric) as well. For example, Fig. 1(b)-(d) shows the flat ground, upward triangle, and downward triangle shaped ground between rows of panels. As we will see later, ground-mounted (i.e., zero elevation) panels on upward triangle-shaped ground pattern (G2) is close to the optimal configuration.

In early mornings and late afternoons, shadows cast by the sun are longer, which can create mutual shading between neighboring rows. Such non-uniform illumination may cause reverse breakdown in series-connected cells. We assume three bypass diodes sub-dividing the panel to minimize the adverse effect of non-uniform illumination [26].

### 2.3. Direct and diffused insolation collection (view factor method):

As the panels are E-W facing, it is straightforward to calculate the angle of incidence (AOI: $\theta^{(F)}$ or $\theta^{(B)}$) between the direct solar beam and the front (or back) face of the panel [16]. Assume that the panels are fixed at an elevation $y_0$ from ground and arranged in array as shown in Fig. 3(a). If the reflection characteristics, efficiency under normally incident direct light, and efficiency under diffused light for front $(R(\theta^{(F)}), \eta^{(F)}, \eta_{\text{diff}}^{(F)})$ and back surfaces are known, we can find the power generated per panel area at height $z$ as follows [16],

$$\hat{I}_{PV(dir)}^{(F)}(z) = [1 - R(\theta^{(F)})]\eta^{(F)} I_b \cos\theta^{(F)}, z > \text{shadow} \quad (1)$$

$$\hat{I}_{PV(diff)}^{(F)}(z) = \eta_{\text{diff}}^{(F)} [I_{\text{diff}} \times F_{dz-sky}] \quad (2)$$

Here, $F_{dz-sky}$ is the view factor from a point $z$ on the panel to the unobstructed part of the sky. The power output contribution from direct and diffused sunlight does not depend on the panel to ground distance, nor the shape or reflectivity of the ground (assuming that the ground is shaped such that it does not cast shadows on the panels).

In general, a view factor $F_{A-B}$ represents the fraction of radiation collected by surface $B$ emitted from surface $A$. The expressions for the view factors used in this paper are straightforward and can be found in prior literature or textbooks [16], [27].

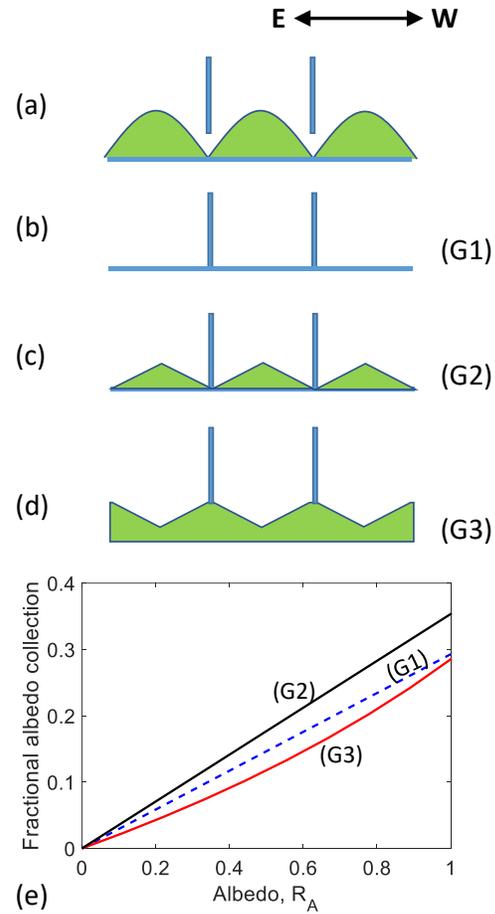

Fig. 1: East-West facing vertical bifacial panel array are shown. (a) shows the general configuration with vertical panel array elevated above arbitrarily shaped periodic ground pattern. (b)-(d) shows the flat ground (G1), upward triangle (G2), and downward triangle (G3) shaped ground between the vertical panels, respectively. (e) Collection of albedo as a fraction of normally incident direct light is compared for the different ground shapes.

## 2.4. Albedo from diffused sunlight (view factor method):

To calculate the albedo collection from the diffused sunlight, we first find the amount of diffused sunlight collected on the sculpted ground under the PV array configuration. Then we can assume the light reflected by the ground as a source of light to be collected by the panels. The amount of diffuse illumination hitting segment $\Delta s$ on the sculpted ground can be written as:

$$I_{\text{Gnd:diff}}(s) = I_{\text{diff}} \times F_{\Delta s-\text{sky}}(s) \quad (3)$$

Here, $F_{\Delta s-\text{sky}}(s)$ is the view factor from the segment $\Delta s$ at position $s$ on the ground to the unobstructed part of the sky. We can now write the corresponding albedo collection on the panels (front face) as follows:

$$\hat{I}^{(F)}_{PV(Alb:diff)}(z) = \eta^{(F)}_{\text{diff}} \sum_s \left[ R_A\, I_{\text{Gnd:diff}}(s) \Delta s\, F^{(F)}_{dz-\Delta s} \right] \quad (4)$$

Here, $F^{(F)}_{dz-\Delta s}$ is the view factor from position $z$ on the panel to the ground segment. The contribution from adjacent periods in the array is small; we therefore only consider the collection of light from ground to panels within the relevant period.

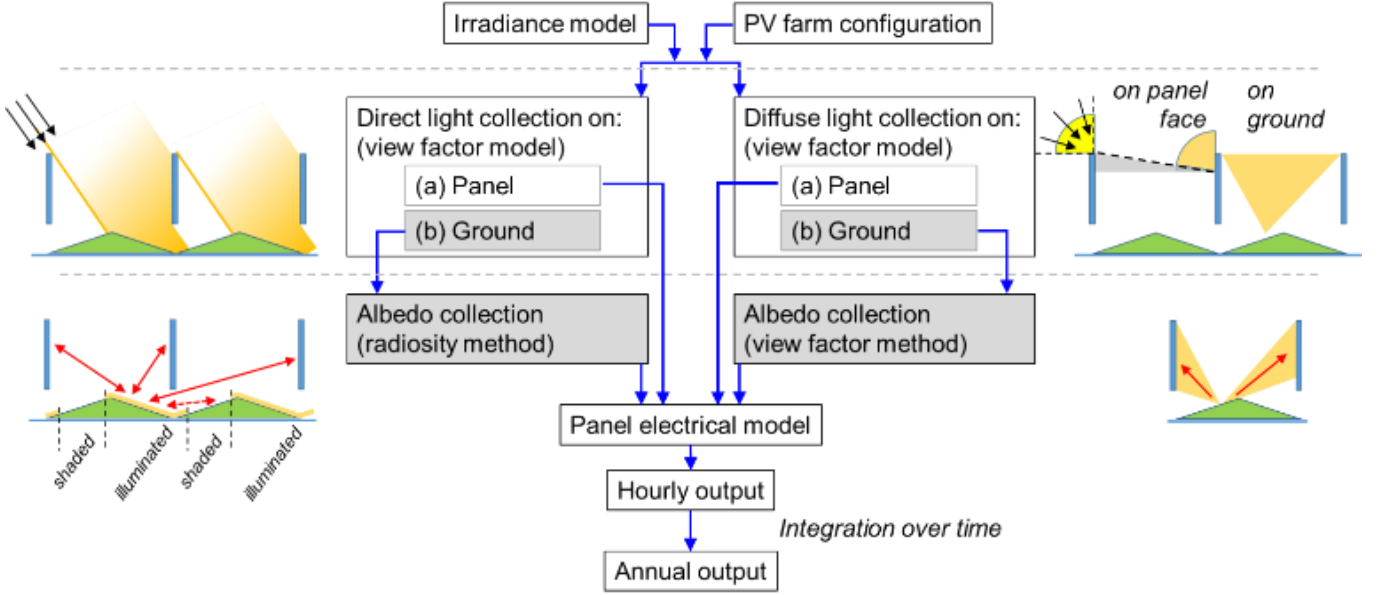

Fig. 2: Calculation flow for determining the output of ground sculpted vertical bifacial PV (GvBF) farm.

## 2.5. Albedo from direct sunlight (radiosity method):

The configuration shown in Fig. 3(a) can in general have any arbitrarily sculpted ground—however, we restrict our analysis to cases when the sculpted ground does not cast any additional shadow on panels. Diffuse reflection from ground can be accounted for using view factor analysis. However, in presence of multiple diffuse surfaces, multiple reflections can occur. These multiple reflections can be accounted for using ray tracing [27]. The radiosity method is used to trace diffuse reflections from multiple surfaces in an enclosure. This technique has its basis in computing heat transfer between diffuse surfaces in an enclosure [27] as well as rendering images in computer graphics [28]. Here, we provide a brief summary of the method. The radiosity is defined as the total energy leaving a surface area with diffuse reflection property. In an enclosure, the radiosity $b_i$ of an element of index $i$ is computed using the following equation:

$$b_i = \sum_j \rho_j b_j F_{ji} + e_i, \quad (5)$$

where $\rho_j$ is the diffuse reflection of each element. In (5), the initial reflection from an element of index $i$ is $e_i = \rho_i H_i$, where $H_i$ is the initial solar irradiation received on the ground element. The view factor between surface elements $F_{ji}$ accounts for multiple reflections between surfaces enclosed.

In an enclosure with $N$ surface elements, (5) can be translated into the matrix equation

$$B = (I - \rho F)^{-1} E, \quad (6)$$

where $B^T = [b_1\ b_2\ \cdots\ b_N]$, $E^T = [e_1\ e_2\ \cdots\ e_N]$. The view factor matrix $F$ contains the mutual view factors between elements involved in the enclosure, with self-view factors $F_{ii} = 0$, since all the elements are assumed to be flat.

It is worth mentioning that the radiosity computation in its form in (6) is computationally expensive in general. However, for the problem in hand, there exist a few number of involved surfaces, each can be decomposed into a finite number of elements. The accuracy of the solution is well-maintained with a minimal number of segments, without the need to use the progressive refinement approach [29].

For the problem of an array of periodic vertical solar panels mounted on or elevated above the ground, we assume that the ground and the panels are infinitely extended in one direction. Thus, the corresponding view factors can be simply computed using the crossed-strings method [27]. A single panel is assigned a single index and zero reflectivity while the ground is decomposed into a number of segments having an albedo reflectivity $R$. Hence, the normalized fractional power received on a single panel due to direct albedo reflection is the first element of the vector $P = FB$. The calculation is performed for both the front and back surfaces of the panel. Hence, the power generated by reflected direct sunlight can be written as:

$$\hat{I}^{(F)}_{PV(Alb:dir)}(z) = \eta^{(F)}_{diff} [I_{dir} \times P(1)]. \quad (7)$$

In principle, elevated panels above the ground surface can receive albedo light reflection from all exposed points on the ground. However, the view factors between a given panel and a ground segment will decrease considerably with the separation distance between them [30]. Hence, a good first order approximation is to consider two periods of the full array. This arrangement is depicted in the bottom left of Fig. 2.

The shadow of the panel on the ground changes with the solar elevation angle. Over the course of the day, only the portions of the ground that are exposed to sunlight are contributing to direct albedo reflection. In some situations, an adjacent panel or a sculpted ground segment can block a portion of the light reflected by the ground, thus reducing the view factor between ground segments and the panel. These situations are computed carefully in the model.

Similar calculations will result in the illumination collection components for the back face as well. All the light collection components by the panel faces are now used to estimate hourly panel output (assuming 3-bypass diode per panel) based on the analytical approach in Ref. [26]. Instead of individual panel output, we focus on the farm yield (i.e., output per land area). The diurnal results are integrated to obtain monthly and annual farm yield.

The model described above can calculate the albedo light collection from arbitrarily curved surface. The curved surface is of course symmetric around the center of the period due to the vertical panel array configuration. Therefore, as a first order estimate, using Hottel's crossed string rule [27], we see that the curved surface can be replaced by two straight lines. For example, a periodic upward hemispherical ground pattern would be equivalent to the upward triangular ground pattern. That is why we will focus on triangular ground shapes for the rest of the paper.

## 3. Physical understanding

There are three questions that we want to answer: (i) Is there an optimum ground pattern? (ii) Given the pattern, what is the optimum period for the panel array? (iii) What improvement, if any, can we expect from the optimized panel array?

Consider the three ground patterns shown in Fig. 1(a)-(c): the flat ground, upward triangle, and downward triangle, labeled G1, G2, and G3, respectively. For a condition when the sun is exactly normal to the ground (e.g., noon at the equator), there is no shading on the ground from the direct light. In such a scenario, the albedo light collection as a fraction of the direct light is shown for various albedo values $R_A$. As expected, for the flat ground G1, the fractional albedo collection increases linearly with $R_A$. At $R_A = 1$, the light collection is limited by the view factors from ground to sky, and panel face to ground. For the downward triangle G3, the fractional albedo collection is worse than G1 for $R_A < 1$. This is because, a part of the light bounces between the ground facing each other—there is of course some loss at each bounce for $R_A < 1$. Finally, the upward triangle G2 is shaped to reflect the light primarily towards the panel faces, and as seen in Fig. 1(d), G2 shows $> 15\%$ increase in fractional albedo collection compared to the trivial flat ground G1.

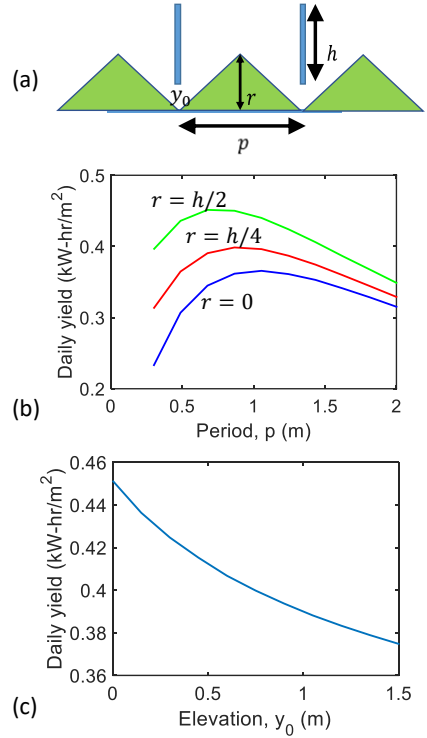

Fig. 3: (a) Vertical bifacial panel array placed elevated $y_0$ from an upward triangular ground. (b) Daily yield (pear land area) of the solar farm (with $y_0 = 0$) as a function of ground triangle height $r$ and array period $p$. (c) Daily yield (per land area) is shown as a function of the panel elevation $y_0$. The values of $r$ and period $p$ are co-optimized at each $y_0$. The calculations are for Washington DC (Sept. 22).

Now that we know we want to choose the up-triangle ground shape for better output, we need to optimize the 'ground sculpted vertical bifacial PV farm' design. Given a specific panel size (width $h = 1$ m), the parameters that define this PV farm are: period of panel array $p$, panel elevation from ground $y_0$, and the height of the ground triangle $r$ as shown in Fig. 3(a).

Fig. 3(b) shows the integrated output of a given day (Sep. 22) in Washington DC for $y_0 = 0$. The line marked $r = 0$ corresponds to the conventional flat ground consistent with our prior work [16]. The daily yield per land area has an optimum period $p$. For small $p$, mutual shading quickly degrades the output. And, at larger $p$, the panels miss a large fraction of light hitting the ground.

Next, as we increase the height $r$ of the triangular ground, essentially tilting the ground towards the panel, thereby steering the albedo light mostly towards the panel faces. As a result, we observe that the output increases as we increase $r$. However, if $r$ is increased beyond (h/2), the ground would cast a shadow on the panels and the output will reduce dramatically. Therefore, for ground mounted panels (i.e., $y_0 = 0$), the triangular ground height can at most be half the panel size, i.e., $r = h/2$—and, as discussed above, this will also result in the highest possible output with optimized period $p$. In general, we would choose $r = y_0 + h/2$ for panels fixed at elevation $y_0$.

Finally, we need to optimize $y_0$. As we increase $y_0$, the 'viewing angle', i.e., view factor between a panel face and the illuminated part of the ground reduces, thereby reducing the albedo collection. Moreover, with increasing $y_0$, larger portion of the ground is exposed to each other. This increases light scattering in between the ground faces (similar to the ground shape G3) and increases scattering loss on the ground. We can therefore conclude that, ground mounting ($y_0 = 0$) the panels are the best choice for this configuration. On the other hand, as seen in literature [6], [5], the standalone tilted bifacial panels collect more albedo light with increased $y_0$. The monotonic decrease in output with $y_0$ is specific to vertical bifacial panel array. Our explanation is also supported by a detailed numerical analysis shown in Fig. 3(c), indicating monotonous decrease in daily yield as $y_0$ is increased. In this plot $(p, r)$-pair is optimally chosen for each value of $y_0$.

To summarize, we can set $y_0 = 0$, $r = h/2$ and then optimize period $p$ to maximize yearly yield. The optimum value of $p$ will depend on the location on earth and the weather (i.e., clearness index).

## 4. Global energy yield

Let us now focus specifically on various locations on earth at latitude 40°N. At a given latitude, sunlight travels through the same thickness of atmosphere (air mass). This results in the same terrestrial insolation under a clear sky assumption. However, as the sky clearness is different at various longitudes, we observe variation in GHI, DNI, and DHI, even at same latitude. We compare the yearly yield of ground sculpted vertical bifacial PV farm (GvBF), conventional vertical bifacial PV farm (vBF), and optimally tilted monofacial PV farm in Fig. 4. We assume an extra 10% loss in output due to soiling for the tilted monofacial panels; in practice, the soiling loss can be considerably higher [31], [32].

The integrated annual yield of the various PV farms (optimized) are shown in Fig. 4(c). In all cases, output increases with annual mean clearness index $k_{TA}$ as the input GHI is high at higher $k_{TA}$. For monofacial PV, we assume 10% output penalty due to higher soiling compared to vertical panels. There is a 15% or more loss in output of vBF with typical ground albedo reflection $R_A = 0.3$ (the earth's average albedo) compared to the monofacial PV farm. Even if the albedo is increased to $R_A = 0.5$ by covering the ground with artificial material, the vBF cannot exceed or match the monofacial farm.

The GvBF (with $R_A = 0.5$) shows exceptional advantage over vBF as well as monofacial farm. As shown in Fig. 4(b), the GvBF shows up to 50% gain in output compared to monofacial farm, especially in cloudy regions (low $k_{TA}$). A closer observation of earth's map and Fig. 4(a) indicates that $k_{TA} > 0.6$ occurs primarily over oceans. Therefore, at all locations of interest (i.e., land), GvBF yields higher output than monofacial farms.

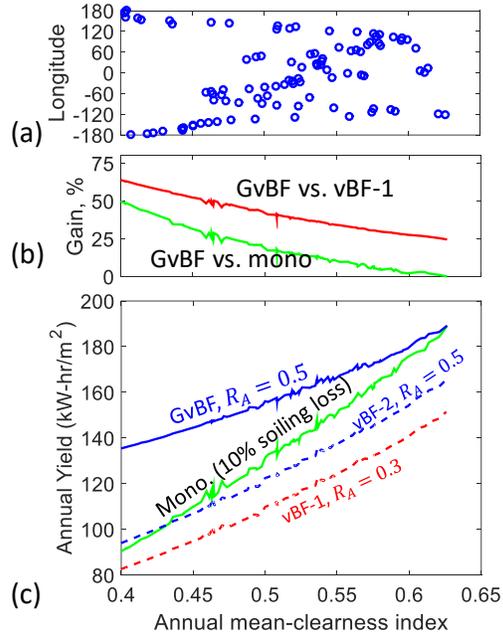

Fig. 4: Annual yield of various monofacial and bifacial PV farm configurations as a function of clearness index (at latitude 40°N) are shown in (c). The vBF solar farms with flat ground and $R_A = 0.3$ and 0.5 are indicated as vBF-1 and vBF-2, respectively. The output gain observed in GvBF compared to other configurations are shown in (b). (a) shows the longitude locations corresponding to the clearness index values at latitude 40°N.

The ground pattern for the GvBF solar farm would be artificially sculpted, and the ground material may not necessarily be natural (grass or sand). Although we choose $R_A = 0.5$ for our artificial material, un-weathered white roofing membranes may have $R_A \sim 0.88$ [23]. With time, this reflectance would decrease; however, $R_A = 0.5$ could still be a conservative estimate with occasional maintenance [23].

5. Summary

In this work, we presented a numerical model for vertical bifacial panel arrays elevated to a specified height upon a strategically patterned ground. The model was applied to study the optimum elevation and ground shape for maximum annual yield. We predict that the optimum condition is achieved when the vertical panels of size $h$ are placed on the ground (no elevation) with the ground sculpted to an upward triangle pattern with height $h/2$. Using a ground material designed to have an effective albedo reflection $R_A = 0.5$, the optimal ground sculpted vertical bifacial PV farm (GvBF) yields more energy than the monofacial farm. In fact, compared to the optimum monofacial farm, the GvBF has 50% more annual energy output in regions with somewhat cloudy sky ($k_{TA} < 0.45$).

Here, we have focused on vertical bifacial panel arrays, which can have much less soiling than tilted monofacial panels. In our analysis, we have assumed an extra 10% soiling loss for the tilted monofacial panels. This value is not absolute for all practical situations—it will vary with the local soiling rate as well as the cleaning cycle—and can be much worse. In the end, the integrated energy output along with bifacial versus monofacial panel costs, and the cleaning costs (integrated over the farm lifetime) will define the difference in the levelized cost of energy (LCOE). In our studies of energy yields, we found that the proposed GvBF configuration will have significant advantages over the conventional monofacial farms, especially, but not only, in moderately to highly-cloudy locations.

**Acknowledgement**

We gratefully acknowledge Dr. Chris Deline from NREL, and Dr. Joshua S. Stein and Dr. Cliff Hansen from Sandia National Laboratories for helpful discussions. This work was partly supported by the National Science Foundation under Grant No. #1724728, and the Solar Energy Research Institute for India and the U.S. (SERIIUS) funded jointly by the U.S. Department of Energy subcontract DE AC36-08G028308 and the Government of India subcontract IUSSTF/JCERDC-SERIIUS/2012. Support was also provided by the National Science Foundation Award EEC 1454315 – CAREER: Thermophotonics for Efficient Harvesting of Waste Heat as Electricity, the Department of Energy, under DOE Cooperative Agreement No. DE-EE0004946 (PVMI Bay Area PV Consortium), and the NCN-NEEDS program under Contract 1227020-EEC.